\documentclass[final]{ws-mpla}
\usepackage[superscript]{cite}
\usepackage{wrapfig}
\newcommand{\be}{\begin{equation}}
\newcommand{\ee}{\end{equation}}
\newcommand{\ba}{\begin{eqnarray}}
\newcommand{\ea}{\end{eqnarray}}
\newcommand{\la}{\langle}
\newcommand{\ra}{\rangle}
\newcommand{\di}{ {\rm d} }
\newcommand{\Todd}[1]{\underline{\underline{#1}}}
\renewcommand{\d}{{\mathrm{d}}}
\renewcommand{\draftnote}{}

\begin{document}

\markboth{Avakian, Efremov, Schweitzer, Teryaev, Yuan, Zavada}
{Non-perturbative aspects of TMDs}

\catchline{}{}{}{}{}

\title{Insights on non-perturbative aspects of TMDs from models}

\author{H.~AVAKIAN$^1$, A.~V.~EFREMOV$^2$, \\ 
	P.~SCHWEITZER$^3$, O.~V.~TERYAEV$^2$, F.~YUAN$^{4,5}$, P.~ZAVADA$^6$}
\address{
  $^1$ 
       Jefferson National Accelerator Facility, Newport News, VA 23606, U.S.A.\\
  $^2$ Bogoliubov Laboratory of Theoretical Physics, JINR, 141980 Dubna, Russia\\
  $^3$ Departement of Physics, University of Connecticut, Storrs, CT 06269, U.S.A.\\
  $^4$ RIKEN BNL Research Center, Building 510A, BNL, Upton, NY 11973, U.S.A.\\
  $^5$ Nuclear Science Division, 
  Berkeley National Laboratory, Berkeley, 
  CA 94720, U.S.A.\\
  $^6$ Institute of Physics, Academy of Sciences of the Czech Republic, 
  Na Slovance 2, CZ-182 21 Prague 8, Czech Republic}

\maketitle

\pub{Received 15 October 2009}{}

\begin{abstract}
  Transverse momentum dependent parton distribution functions 
  are a key ingredient in the description of spin and azimuthal
  asymmetries in deep-inelastic scattering processes.
  Recent results from non-perturbative calculations in effective
  approaches are reviewed, with focus on relations among different 
  parton distribution functions in QCD and models.

\end{abstract}

\ccode{13.88.+e, 
      13.85.Ni,  
      13.60.-r,  
      13.85.Qk}  

\section{Introduction}
\label{Sec-1:introduction}

TMDs contain so far unexplored information on the nucleon structure
\cite{Collins:2003fm,Collins:2007ph,Collins:1999dz,Hautmann:2007uw}, 
and owing to factorization \cite{Collins:1981uk,Ji:2004wu}
are accessible in leading-twist observables in deeply inelastic processes
\cite{Cahn:1978se,Collins:1984kg,Sivers:1989cc,Efremov:1992pe,Collins:1992kk,Collins:1993kq,Kotzinian:1994dv,Mulders:1995dh,Boer:1997nt,Boer:1997mf,Boer:1999mm,Bacchetta:1999kz,Collins:2002kn,Belitsky:2002sm,Goeke:2005hb,Bacchetta:2006tn,Arnold:2008kf},
on which first data are available, such as
SIDIS \cite{Arneodo:1986cf,Airapetian:1999tv,Avakian:2003pk,Airapetian:2004tw,Alexakhin:2005iw,Diefenthaler:2005gx,Ageev:2006da,Avakian:2005ps,Kotzinian:2007uv,Osipenko:2008rv,Giordano:2009hi,Gohn:2009}
hadron production in $e^+e^-$ annihilations \cite{Abe:2005zx,Ogawa:2006bm,Seidl:2008xc},
Drell-Yan process \cite{Falciano:1986wk,Guanziroli:1987rp,Conway:1989fs,Zhu:2006gx}.
In order to be sensitive to ``intrinsic'' transverse parton momenta it 
is necessary to measure adequate transverse momenta in the final state, 
for example, in SIDIS the transverse momenta $P_{h\perp}$ of produced 
hadrons with respect to the virtual photon momentum.

The SIDIS process is characterized by 18 structure functions:
eight leading, eight subleading and two subsubleading in $1/Q$.
In Bjorken-limit for $P_{h\perp}\ll Q$, which denotes the virtuality of the 
exchanged photon, the eight leading-twist structure functions are described in
one-to-one correspondence in terms of eight leading-twist TMDs 
(and two leading-twist fragmentation functions) \cite{Boer:1997nt}.
Two leading twist azimuthal SSAs 
measured \cite{Airapetian:2004tw,Alexakhin:2005iw,Diefenthaler:2005gx,Ageev:2006da}
in SIDIS off transversely polarized targets, $A_{UT}^{\sin(\phi\pm\phi_S)}$,
which received much attention, already give rise to a first rough 
picture of two novel functions: transversity and Sivers function 
\cite{Efremov:2004tp,Collins:2005ie,Anselmino:2005nn,Vogelsang:2005cs,Efremov:2006qm,Arnold:2008ap,Anselmino:2008sga}.

However, one should not forget that the excitement about SSAs in SIDIS was
initiated by observations of seizable longitudinal (target, beam) SSAs
$A_{UL}^{\sin\phi}$, $A_{LU}^{\sin\phi}$ which are {\sl subleading} in $1/Q$
\cite{Airapetian:1999tv,Avakian:2003pk} and much more precise data are
under way \cite{Gohn:2009}.
In spite of numerous efforts 
\cite{DeSanctis:2000fh,Anselmino:2000mb,Efremov:2001cz,Ma:2002ns,Bacchetta:2002tk,Yuan:2003gu,Gamberg:2003pz,Bacchetta:2004zf,Metz:2004je,Metz:2004ya,Afanasev:2006gw}
there is presently no fully satisfactory explanation, 
what gives rise to these SSAs. 
In the case of the azimuthal asymmetry in unpolarized SIDIS 
$A_{UU}^{\cos\phi}$ on which final \cite{Arneodo:1986cf,Osipenko:2008rv}
and preliminary \cite{Giordano:2009hi} data are available,
the situation is similar.
The interpretation of subleading twist observables is more involved,
because ({\sl assuming factorization}) the 8 subleading in $1/Q$ 
structure functions receive contributions from 16(!) twist-3 TMDs 
(and further twist-3 fragmentation functions) \cite{Goeke:2005hb,Bacchetta:2006tn}.
Similar challenges will be faced when interpreting Drell-Yan data
\cite{Arnold:2008kf}.

In this situation information from models is valuable. 
But often models provide insights on TMDs at low hadronic scales
\cite{Yuan:2003wk,Afanasev:2006gw,Gamberg:2006ru,Gamberg:2007wm,
Jakob:1997wg,Pasquini:2008ax,Efremov:2009ze,Bacchetta:2008af,She:2009jq,
Courtoy:2008dn,work-in-progress,Meissner:2007rx,Avakian:2008dz},
and it is difficult to make reliable estimates for SSAs
at experimentally relevant scales \cite{Boffi:2009sh}.
A particularly elegant way of using models consists therefore 
in exploring relations among TMDs in models. 

In QCD there can be no {\sl exact} but at best {\sl approximate}
relations among quark TMDs, see Sec.~\ref{Sec-2:relations-in-QCD}. 
In relativistic quark models, however, Lorentz-invariance implies that
certain relations among TMDs must exist, see Sec.~\ref{Sec-3:LIRs-in-models}.
Of course, depending on the model further relations among TMDs may also exist,
as we will review in Sec.~\ref{Sec-4:relations-in-models}.
We present first results from bag model,
see Sec.~\ref{Sec-5:relations}, before we conclude in
Sec.~\ref{Sec-5:conlcusions}.

\section{Relations among TMDs in QCD}
\label{Sec-2:relations-in-QCD}

TMDs are defined in terms of light-front correlators with a process-dependent 
gauge-link ${\cal W}$ \cite{Collins:2002kn,Belitsky:2002sm}
(we do not indicate the scale and flavor dependence for brevity) 
\be\label{Eq:correlator}
    \phi(x,\vec{p}_T)_{ij} = \int\frac{\di z^-\di^2\vec{z}_T}{(2\pi)^3}\;e^{ipz}\;
    \la N(P,S)|\bar\psi_j(0)\,{\cal W}\,\psi_i(z)|N(P,S)\ra
    \biggl|_{z^+=0,\,p^+ = xP^+} 
    \ee
by taking traces of (\ref{Eq:correlator}) with $\gamma$-matrices.
This yields linear combinations of TMDs
(the T-odd distributions are $\Todd{\mbox{un}}$derlined)
\ba
\mbox{twist-2:}\;\;
\gamma^+,\; \gamma^+\gamma_5,\; i\sigma^{j+}\gamma_5\;\; 
&\to& 	\hspace{1mm}
	(f_1,\,\Todd{f_{1T}^\perp}),\; 
	(g_1,\,g_{1T}^\perp),\; 
	(h_1,\,h_{1L}^\perp ,\,h_{1T}^\perp,\,\Todd{h_1^\perp})\phantom{\frac11}\\
\mbox{twist-3:}\;\;\,\,
	\gamma^j,\; 
	\gamma^j\gamma_5,\; 
	i\sigma^{+-}\gamma_5 \;\;
&\to&   \hspace{1mm}
	(f^\perp,\,\Todd{f_L^\perp}),\; 
	(g_T,\, g_L^\perp,\, g_T^\perp,\, \Todd{g_{}^\perp}),\; 
	\hspace{3mm} (h_L,\, h_T)\phantom{\frac11} \nonumber\\
	1,\;\gamma_5,\;i\sigma^{\,jk\,}\gamma_5\;\;\;
&\to&	\hspace{18mm}
	(e, \, \Todd{e_T^\perp}), \;
	(\Todd{e_L^{}}, \, \Todd{e_T^{}}), \;\hspace{3mm}
	(\,\Todd{h_{}^{}},\,\,h_T^\perp)\phantom{\frac11}
\ea
where space-indices $j,k$ refer to the plane transverse 
with respect to the light-cone. The $\gamma$-structures not 
listed above give rise to twist-4 objects \cite{Goeke:2005hb}.

It was shown that there are 32 (twist-2, 3, 4)
quark TMDs \cite{Goeke:2005hb} and that the unintegrated correlator,
i.e.\ (\ref{Eq:correlator}) without constraints on $z^+$ and $p^+$,  
contains 32 independent Lorentz structures: 12 Lorentz-scalar 
amplitudes $A_i$ accompanying Lorentz-structures constructed from the
nucleon momentum $P$, nucleon spin $S$, and quark momentum $p$;
and 20 amplitudes $B_i$ which arise by considering that
the Wilson link ${\cal W}$ provides a further Lorentz-vector
$n_-$ characterizing the light-cone direction
\cite{Goeke:2003az}.
Thus, all TMDs are independent, there are no relations among them
\cite{Goeke:2005hb}.

In first works  the amplitudes $B_i$ were not noted 
\cite{Mulders:1995dh,Boer:1997nt}. This implied more TMDs 
than amplitudes, giving rise to so-called 'Lorentz-invariance 
relations' (LIRs). It was later recognised that LIRs are not 
valid in QCD \cite{Kundu:2001pk,Goeke:2003az}.
However, some LIRs hold in a 'Wandzura-Wilczek-(WW)-type-approximation'
\cite{Metz:2008ib,Teckentrup:2009xyz}, see also App.~A.

\begin{wrapfigure}[14]{R}{5.4cm}
\vspace{-5mm}
    \includegraphics[width=5cm]{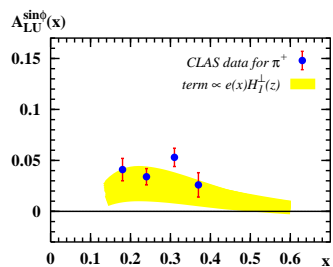}
\vspace{-2mm}
\caption{\label{Fig03:beam-SSA}
    The SSA $A_{LU}^{\sin\phi}$ vs.\ $x$.
    The data are from CLAS \cite{Avakian:2003pk}, and
    the shaded area is the contribution due to
    the chirally odd twist-3 distribution function  \cite{Schweitzer:2003uy}
    $e^q(x)$ and the Collins fragmentation function \cite{Efremov:2006qm}
    $H_1^\perp$.}
\end{wrapfigure}

What WW-type-approximations have in common with the classic 
WW-approximation, 
$g_T(x)\approx\int_x^1\di x^\prime g_1(x^\prime)/x^\prime$
\cite{Wandzura:1977qf},
is that in both cases QCD equations of motion are explored, 
and quark-gluon correlators and current quark 
mass terms neglected, though the nature of neglected operators 
is different.
The classic WW-approximation \cite{Wandzura:1977qf}
holds in parton model
\cite{Jackson:1989ph,Zavada:2002uz,D'Alesio:2009kv},
is supported by theory \cite{Balla:1997hf}, 
and by data with (15--40)$\,\%$ accuracy or better \cite{Accardi:2009au}.

WW-type-approximations were used in literature
\cite{Kotzinian:2006dw,Avakian:2007mv} and, if satisfied within 
a reasonable accuracy, could be helpful for analyses 
of first data \cite{Avakian:2007mv,Metz:2008ib}.
Interestingly, the beam SSA $A_{LU}^{\sin\phi}$ is due to
such quark-gluon (and mass) terms only. The effect is not large, 
as expected in WW-type-approximation \cite{Metz:2008ib},
but clearly non-zero \cite{Avakian:2003pk,Gohn:2009},
see Fig.~\ref{Fig03:beam-SSA}, providing direct insights 
into the physics of quark-gluon correlators.

\section{\boldmath Model-independent relations among TMDs in quark models}
\label{Sec-3:LIRs-in-models}

In the following 'quark model' refers to an effective approach with 
quark (sometimes also antiquark) degrees of freedom but no gluons.
In such models there are no T-odd TMDs, but one can describe 
T-even TMDs. Since in such models there is also no gauge-link, 
it means that there are no $B_i$ amplitudes, see 
Sec.~\ref{Sec-2:relations-in-QCD}. 

Consequently, the LIRs discussed above must be valid
{\sl in any quark model} which respects Lorentz-invariance,
but has no gauge-field degrees of freedom.
There are 5 LIRs 
among the 14 T-even (twist-2 and 3) TMDs, namely \cite{Teckentrup:2009xyz}
\ba
\label{eq:LIR1} g_T(x) \; &\stackrel{\rm LIR}{=}& \; 
	        g_1(x) + \frac{\d}{\d x} g^{\perp(1)}_{1T}(x)\, ,\\ 
\label{eq:LIR2} h_L(x) \; &\stackrel{\rm LIR}{=}& \; 
	      	h_1(x) - \frac{\d}{\d x} h^{\perp(1)}_{1L}(x) \, , \\
\label{eq:LIR4} h_T(x) \; &\stackrel{\rm LIR}{=}& \; 
		- \frac{\d}{\d x} h^{\perp(1)}_{1T}(x) \, , \\
\label{eq:LIR3} g_L^\perp(x) + \frac{\d}{\d x} g_T^{\perp(1)}(x) \; 
		&\stackrel{\rm LIR}{=}& \; 0 \, ,\\ 
\label{eq:LIR5} h_T(x,p_T)-h_T^\perp(x,p_T) \; &\stackrel{\rm LIR}{=}& 
		\; h^{\perp}_{1L}(x,p_T) \, .
\ea
The transverse moments 
$g^{\perp(1)}_{1T}(x)=\!\int\di^2p_T\frac{p_T^2}{2M^2}g^{\perp}_{1T}(x,p_T)$, etc.\
are well-defined in models (presuming adequate regularization, when necessary).

\section{\boldmath Relations among TMDs in quark models}
\label{Sec-4:relations-in-models}

If in a relativistic quark model all 9 T-even amplitudes $A_i$ in the 
Lorentz-expansion of the correlator (\ref{Eq:correlator}) happen to be 
different, this is the end of the story. Then, besides the {\sl quark model 
relations} (\ref{eq:LIR1})--(\ref{eq:LIR5}) there are no further relations 
among TMDs. However, several noteworthy model relations 
among TMDs were found, which are supported in many 
\cite{Jakob:1997wg,Avakian:2008dz,Pasquini:2008ax,Efremov:2009ze,She:2009jq,Meissner:2007rx,Bacchetta:2008af}
though not all \cite{Bacchetta:2008af} quark models.
These model relations, unlike (\ref{eq:LIR1})--(\ref{eq:LIR5}) are not 
required to hold in a quark model because of Lorentz-invariance or any other
(apparent) symmetry. The fact they are supported in very different approaches
\cite{Jakob:1997wg,Avakian:2008dz,Pasquini:2008ax,Efremov:2009ze,She:2009jq,Meissner:2007rx,Bacchetta:2008af}
makes such relations interesting. Let us discuss three examples.

(i). The first example is the relation first observed in the spectator model 
\cite{Jakob:1997wg}
\be\label{Eq:rel-IV}
  g_{1T}^{\perp q}(x,p_T) \;\,\stackrel{\rm model}{=}\;\,  
  -\, h_{1L}^{\perp q}(x,p_T)\;.\phantom{\frac11}
\ee
This relation has an appealing partonic interpretation: the 'polarization
of transversely polarized quarks in a longitudinally polarized nucleon'
is opposite to the  'polarization of longitudinally polarized quarks in a 
transversely polarized nucleon'.
Subtracting the LIRs (\ref{eq:LIR1}) and (\ref{eq:LIR2}) from each other, 
which hold in any consistent quark model, see
Sec.~\ref{Sec-3:LIRs-in-models}, and using (\ref{Eq:rel-IV}) 
we arrive at the interesting conclusion that 
$g_1^q(x)-h_1^q(x)=g_T^q(x)-h_L^q(x)$.
This means that in the class of models supporting (\ref{Eq:rel-IV}) 
the difference of the twist-3 distributions $g_T$ and $h_L$
is a 'measure relativistic effects' in the nucleon
\cite{Jaffe:1991ra} as is the difference of $g_1$ and $h_1$.

(ii). The second example, the 'pretzelosity' relation 
first found in bag model \cite{Avakian:2008dz}, makes the last statement
even more quantitative, namely 
\be\label{Eq:rel-VI}
g_1^q(x,p_T) - h_1^q(x,p_T) \;\,\stackrel{\rm model}{=}\;\, 
h_{1T}^{\perp(1)q}(x,p_T)\;,\phantom{\frac11}
\ee
i.e.\ pretzelosity's (unintegrated) transverse moment 
$h_{1T}^{\perp(1)}(x,p_T)=\frac{p_T^2}{2M^2}h^{\perp}_{1T}(x,p_T)$
is just this 'measure of relativistic effects.'
What makes this function furthermore interesting is the fact
that it seems, at least in models, to be most directly related to
quark orbital momentum \cite{She:2009jq} and moreover 'measures'
also the non-sphericity of the transversal spin distribution in
the nucleon \cite{Miller:2007ae}.

(iii). The third example is the remarkable {\sl non-linear relation}
observed in the covariant parton model \cite{Efremov:2009ze}. It 
connects all T-even, chirally odd, leading twist TMDs,
\be\label{Eq:rel-non-linear}
      \frac{1}{2}\,\biggl[h_{1L}^{\perp q}(x,p_T)\biggr]^2 
      \;\,\stackrel{\rm model}{=}\;\, 
      - \,h_1^q(x,p_T)\,h_{1T}^{\perp q}(x,p_T)\;.
\ee
It can be used to make a particularly robust prediction.
From (\ref{Eq:rel-non-linear}) it follows that $h_1^q$ 
and $h_{1T}^{\perp q}$ must have opposite signs. Since on proton 
$A_{UT}^{\sin(\phi-\phi_S)}(\pi^+)\propto h_1 \otimes H_1^\perp$
is found positive \cite{Airapetian:2004tw}, we conclude that  
$A_{UT}^{\sin(\phi-3\phi_S)}(\pi^+)\propto h_{1T}^\perp \otimes H_1^\perp$
should be negative. For $\pi^-$ the situation is opposite.
It will be interesting to see whether this prediction will be
confirmed in experiment. $A_{UT}^{\sin(\phi-3\phi_S)}$ could be measured
at Jefferson Lab \cite{Avakian:2008dz}.

\section{Relations among TMDs in bag model}
\label{Sec-5:relations}

A question which emerges, when reviewing the quark model 
relations in Sec.~\ref{Sec-4:relations-in-models}, is: 
{\sl how many relations among TMDs exist in quark models?}

It would be interesting to formulate the general 
conditions which must be satisfied in a model, such that relations 
of the kind (\ref{Eq:rel-IV})--(\ref{Eq:rel-non-linear}) hold.
Before attempting this, however, it is instructive to consider
a specific quark model, and study all (T-even) TMDs in this framework.
For definiteness, we shall use here the bag model \cite{work-in-progress}.

%
\begin{figure}[t!]
             \includegraphics[width=4.5cm]{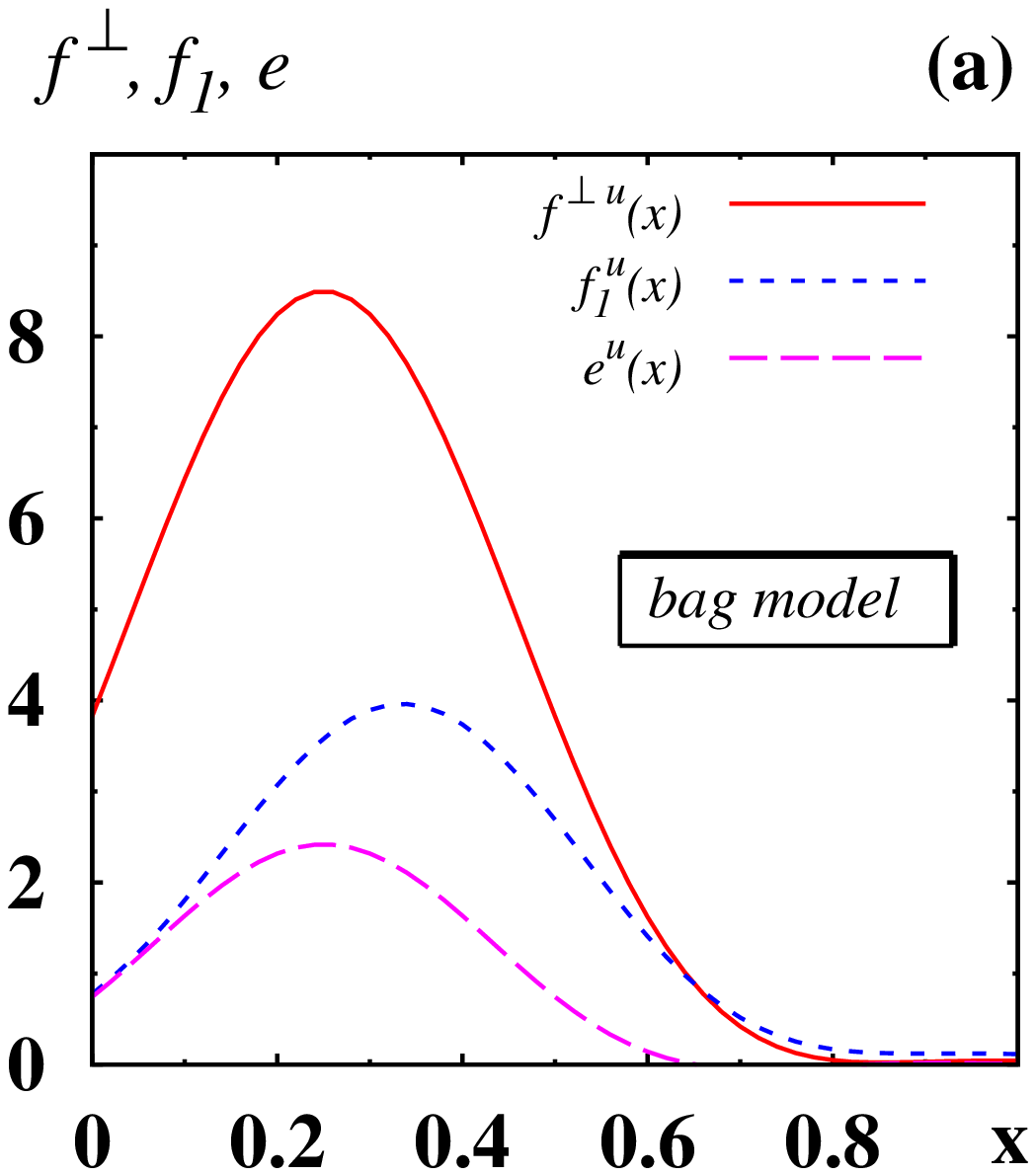}
\hspace{-5mm}\includegraphics[width=4.5cm]{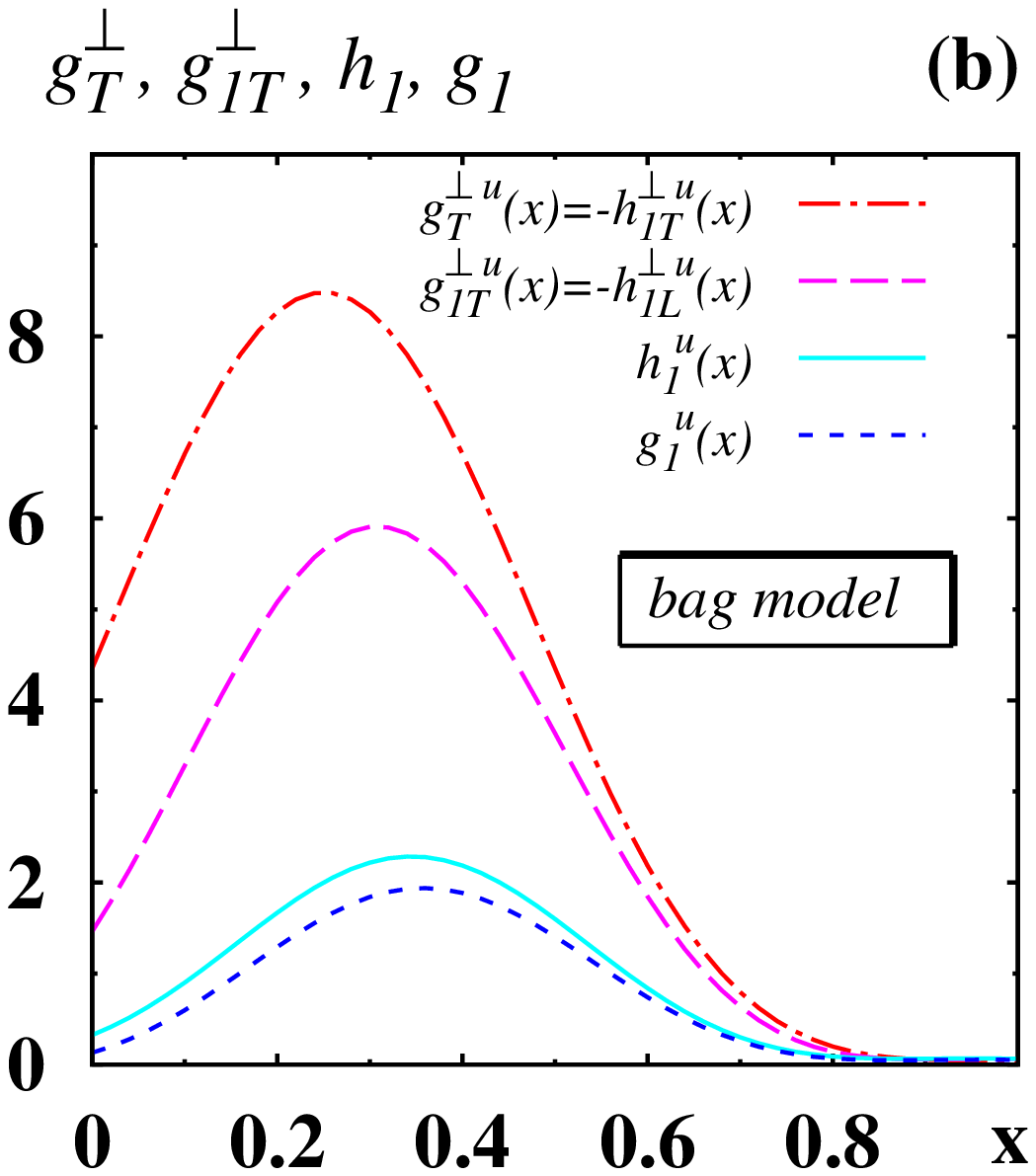}
\hspace{-5mm}\includegraphics[width=4.5cm]{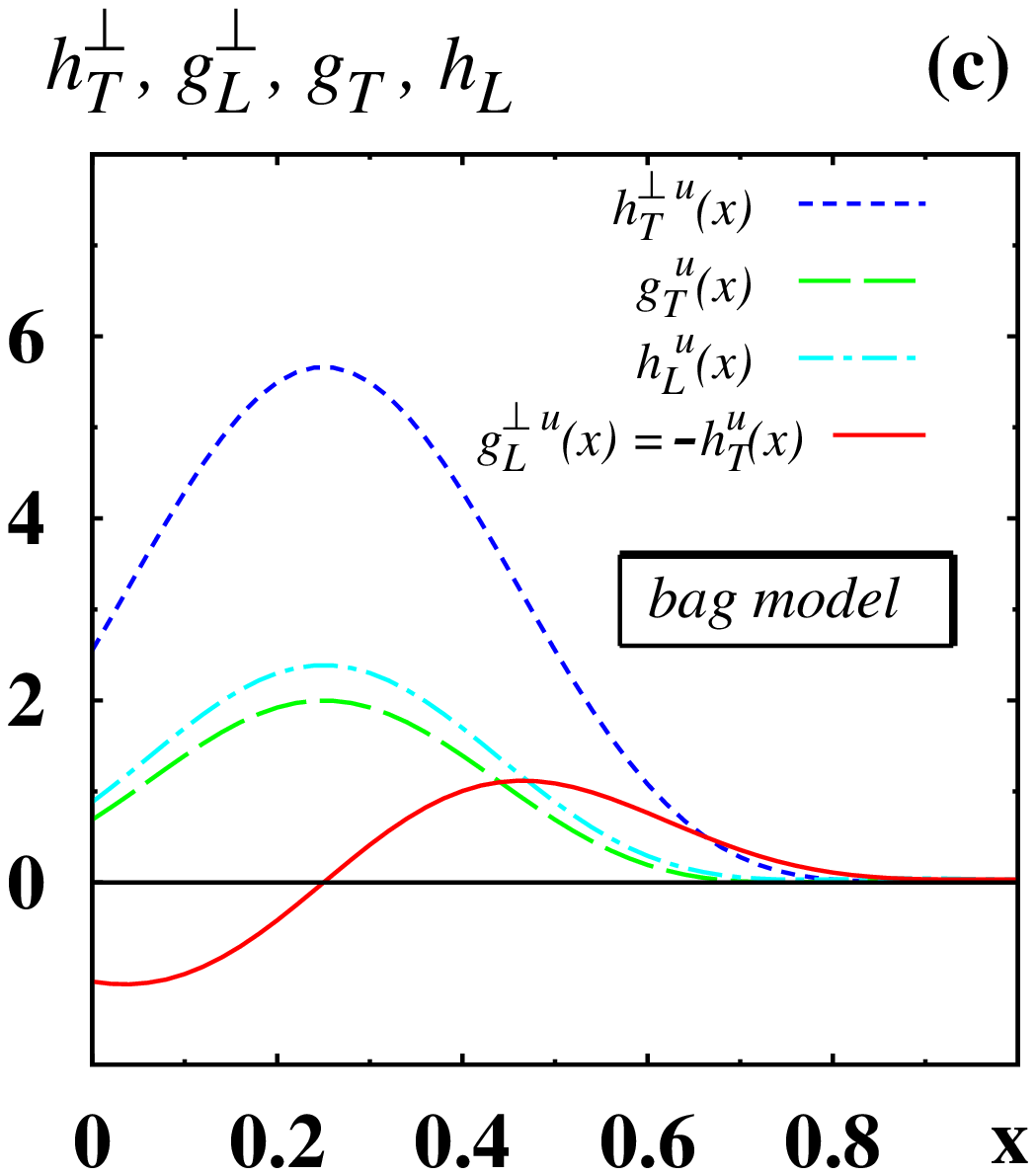}
\caption{\label{Fig-1:TMDs-bag}
    (a)
    The unpolarized functions $f^{\perp u}(x)$, $f_1^u(x)$, $e^u(x)$ vs.\ $x$
    from the bag model at low~scale. 
    (b)
    The polarized functions $g_T^{\perp u}(x)=-h_{1T}^{\perp u}(x)$, 
    $g_{1T}^{\perp u}(x)=-h_{1L}^{\perp u}(x)$, $h_1^u(x)$, $g_1^u(x)$ 
    plotted vs.\ $x$.
    (c)
    The polarized functions $h_T^{\perp u}(x)$, 
    $g_L^{\perp u}(x)=-h_T^u(x)$, $g_T^u(x)$, $h_L^u(x)$ as functions of $x$.
    Distribution functions of $d$-quarks are related to those of $u$-quarks
    by SU(6) spin-flavour symmetry.}
\end{figure}
%

Fig.~\ref{Fig-1:TMDs-bag} shows TMDs in the bag \cite{work-in-progress}.
Here we focus on relations. In the bag model, there are 9 linear relations 
among the 14 (twist-2, 3) T-even TMDs, namely:
\ba
    {\cal D}^q\,f_1^q(x,p_T) + g_1^q(x,p_T) 
    \;\,&\stackrel{\rm bag}{=}&\;\, \;2 h_1^q(x,p_T)
    \label{Eq:rel-I} \\
    {\cal D}^q\,e^q(x,p_T) + h_L^q(x,p_T) 
    \;\,&\stackrel{\rm bag}{=}&\;\, 2g_T^q(x,p_T)
    \label{Eq:rel-II}\\
    {\cal D}^q\,f^{\perp q}(x,p_T)
    \;\,&\stackrel{\rm bag}{=}&\;\, h_T^{\perp q}(x,p_T)
    \label{Eq:rel-III}\\
    \hspace{5mm}g_{1T}^{\perp q}(x,p_T) 
     \;\,&\stackrel{\rm bag}{=}&\;\, 
     -\, h_{1L}^{\perp q}(x,p_T)\;\label{Eq:rel-IVa}\\
    \hspace{5mm}g_T   ^{\perp q}(x,p_T)    
    \;\,&\stackrel{\rm bag}{=}&\;\, 
    -\, h_{1T}^{\perp q}(x,p_T)\label{Eq:rel-V}\\
    \hspace{5mm}g_L   ^{\perp q}(x,p_T)
    \;\,&\stackrel{\rm bag}{=}&\;\, 
    -\, h_T   ^{      q}(x,p_T)\label{Eq:rel-VIa}\\
    \hspace{5mm}g_1^q(x,p_T) - h_1^q(x,p_T) 
    \;\,&\stackrel{\rm bag}{=}&\;\, h_{1T}^{\perp(1)q}(x,p_T)\;,
    \label{Eq:measure-of-relativity}\\
    \hspace{5mm}g_T^{      q}(x,p_T)-h_L^{      q}(x,p_T) 
    \;\,&\stackrel{\rm bag}{=}&\;\,  
    h_{1T}^{\perp(1)q}(x,p_T)\label{Eq:rel-VIII}\\
    \hspace{5mm}h_T^{      q}(x,p_T)-h_T^{\perp q}(x,p_T) 
    \;\,&\stackrel{\rm bag}{=}&\;\,  
    h_{1L}^{\perp q}(x,p_T)\label{Eq:rel-IX}
\ea
where ${\cal D}^q=\frac{P_q}{N_q}$ with $P_u=\frac43$,
$P_d=-\frac13$, $N_u=2$, $N_d=1$ as dictated by SU(6) spin-flavor
symmetry. It can be shown that there are no further linear relations
in the bag \cite{work-in-progress}.
Of course, Eqs.~(\ref{Eq:rel-I}--\ref{Eq:rel-IX}) represent 
just one way of writing down these relations. 

Some comments are in order. Some of the relations were discussed
previously in literature.
Relation (\ref{Eq:rel-I}) was discussed in its $p_T$-integrated 
\cite{Jaffe:1991ra} and unintegrated version \cite{Pasquini:2008ax}, 
and (\ref{Eq:rel-II}) in its integrated version \cite{Signal:1997ct}.
Eqs.~(\ref{Eq:rel-IVa}), (\ref{Eq:measure-of-relativity})
were reviewed in Sec.~\ref{Sec-4:relations-in-models}, and
(\ref{Eq:rel-IX}) is a LIR. Of course, also all other T-even LIRs 
hold in the bag, as it should be, see Sec.~\ref{Sec-3:LIRs-in-models}.
Relations (\ref{Eq:rel-III}), (\ref{Eq:rel-V}), (\ref{Eq:rel-VIII}) 
were not mentioned previously in literature, but 
the latter 2 hold in the spectator model \cite{Jakob:1997wg}.

The relations (\ref{Eq:rel-I})--(\ref{Eq:rel-III}) connect polarized  
and unpolarized TMDs. SU(6) symmetry is not sufficient for them to hold 
\cite{Jakob:1997wg}. Apparently, these relations require strong
model assumptions. 
From the point of view of model dependence, it is 'safer' 
\cite{Avakian:2008dz}
to compare relations which include only polarized TMDs. Relations
(\ref{Eq:rel-IVa})--(\ref{Eq:rel-IX}) are of this type, 
and indeed we find them supported in a larger class of quark models.

\begin{wrapfigure}[16]{R}{5.2cm}
\vspace{-6mm}
    \includegraphics[width=5cm]{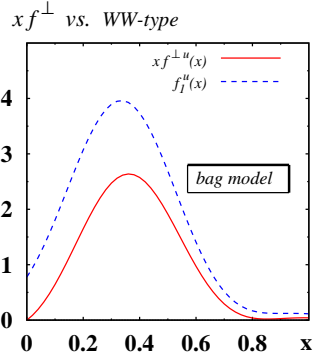}
\vspace{-2mm}
\caption{\label{Fig02:WW-type-approx}
    $xf^{\perp u}(x)$ and $f_1^u(x)$ vs.\ $x$.
    These functions would be equal in a WW-type
    approximation. }
\end{wrapfigure}

Another worthwhile commenting feature is that the TMDs
$g_L^{\perp u}(x)=-h_T^u(x)$ are the only ones,
which have a zero in the valence-$x$ region. 
Actually, this feature is not surprising due to the LIRs 
(\ref{eq:LIR4}) and (\ref{eq:LIR3})
but expected to hold in all quark models where
$h^{\perp(1)}_{1T}(x)$ and $g_T^{\perp(1)}(x)$ have valence-like 
shapes with extrema in valence-$x$ region, as is the case in the 
bag model.

The bag model allows to test WW-type approximations, 
see Sec.~\ref{Sec-2:relations-in-QCD}. From QCD equations of 
motion we have $xf^{\perp q}(x)=x\tilde{f}^{\perp q}(x)+f_1^q(x)$.
In that model the quarks are not free particles, but confined
by the bag. Hence, one cannot expect the pure-interaction-dependent
piece $\tilde{f}^{\perp q}(x)$ to be zero.
The interesting question is whether it is small or large.

Fig.~\ref{Fig02:WW-type-approx}b shows $xf^{\perp u}(x)$ in comparison to 
$f_1^u(x)$. 
The bag model roughly supports this WW-type approximation at the low scale.
It remains to be seen to which extent the bag, which is a 
model for confinement, is capable to 'mimic gluons' \ \cite{Jaffe:1991ra} 
and to simulate the QCD interaction-dependent terms.

In the context of $f^{\perp q}(x)$ such an approximation plays an
important role in the interpretation \cite{Anselmino:2005nn} of the EMC 
data on the azimuthal asymmetry $A_{UU}^{\cos\phi}$ in unpolarized SIDIS 
\cite{Arneodo:1986cf} as being due to the Cahn effect \cite{Cahn:1978se}.

\section{Conclusions}
\label{Sec-5:conlcusions}

We reviewed recent results from quark models on TMDs, with 
particular emphasis on model relations among TMDs. Interestingly,
there is a class of relations ('LIRs') which any consistent
relativistic quark model must satisfy.
The dynamics in quark models certainly oversimplifies QCD. Nevertheless 
such approaches catch important features of the nucleon properties.
Whether this is also the case with the (LIRs $=$) {\sl quark model relations}
(\ref{eq:LIR1})--(\ref{eq:LIR5}) will be clarified by data.

We also reviewed that in various models further relations among
TMDs have been found, which are not related to any (apparent) symmetry.
We sketched how such relations can be used to make robust predictions,
robust in the sense that they are supported by a large class of models.
We presented first results from a complete study of T-even (twist-2 and 3)
TMDs in a bag model.

Although such quark model relations are appealing, especially in the 
context of pretzelosity \cite{Avakian:2008dz} promisingly related to 
quark orbital momentum \cite{She:2009jq}, one should keep in mind
that they break down in models with gauge-field degrees of freedom
\cite{Meissner:2007rx}. Of course, in QCD all TMDs 
are independent, and no relations exist.

At the same time, it is worth to recall that quark models 
have a long history of successful applications in phenomenology.
In spite of all their limitations, these models do catch some
key features of nucleon properties. This encourages to explore 
quark \cite{Boffi:2009sh} or parton \cite{Efremov:2004tz} model
approaches also in the context of TMDs.

Eventually, data will show to which extent such quark model 
relations hold in nature. Until then models remain the
important tool to skill our understanding of non-perturbative
properties of TMDs.

\section*{Acknowledgments}

This paper is dedicated to Anatoli V. Efremov on the occasion of his 75th 
birthday by the other co-authors. Several of us are grateful to the organizers
of the workshop “Recent Advances in Perturbative QCD and Hadronic Physics”, ECT*,
Trento (Italy) for the efforts which made this memorable event possible. 

\vspace{3mm}

\noindent
A.~E.\ and O.~T. are suppoted by the Grants RFBR 09-02-01149 and 07-02-91557, 
RF MSE RNP.2.2.2.2.6546 (MIREA) and by the Heisenberg-Landau Program of JINR.
The work was supported in part by DOE contract DE-AC05-06OR23177, under
which Jefferson Science Associates, LLC,  operates the Jefferson Lab.
F.~Y.~is grateful to RIKEN, Brookhaven National Laboratory and the U.S.\
Department of Energy (contract number DE-AC02-98CH10886) for providing
the facilities essential for the completion of this work.

\appendix

\section{The earlier analogs of LIRs} 
 
The earlier analogs of LIRs are the relations at 
twist-3 level \cite{Efremov:1983eb}
providing the independence of {\it physical observables} on
the specific choice of the light-cone vector $n_-$.
These relations lead, in particular, to
the Burkhardt-Cottingham sum rule and, after neglecting the 
quark-gluon correlations \cite{Teryaev:1995um,Anikin:2001ge},
to WW-relations. 

Recently, $n_-$-independence  was generalized
for the case of twist 3 contributions to exclusive impact factors
\cite{Anikin:2009bf}.

\newpage


\begin{thebibliography}{99}

\bibitem{Collins:2003fm}
  J.~C.~Collins,
  Acta Phys.\ Polon.\ B {\bf 34} (2003) 3103.

\bibitem{Collins:2007ph}
  J.~C.~Collins, T.~C.~Rogers and A.~M.~Stasto,
  Phys.\ Rev.\  D {\bf 77} (2008) 085009.

\bibitem{Collins:1999dz}
  J.~C.~Collins, F.~Hautmann,
  Phys.\ Lett.\  B {\bf 472} (2000) 129,
  JHEP {\bf 0103} (2001) 016. 

\bibitem{Hautmann:2007uw}
  F.~Hautmann,
  Phys.\ Lett.\  B {\bf 655} (2007) 26. 

\bibitem{Collins:1981uk}
  J.~C.~Collins and D.~E. Soper,
  Nucl.\ Phys.\ B {\bf 193} (1981) 381,
  and {\bf 213}  (1983) 545E.

\bibitem{Ji:2004wu}
  X.~D.~Ji, J.~P.~Ma and F.~Yuan,
  Phys.\ Rev.\ D {\bf 71} (2005) 034005,
  Phys.\ Lett.\ B {\bf 597} (2204) 299.
  J.~C.~Collins and A.~Metz,
  Phys.\ Rev.\ Lett.\  {\bf 93} (2004) 252001. 

\bibitem{Cahn:1978se}
  R.~N.~Cahn,
  Phys.\ Lett.\ B {\bf 78} (1978) 269.

\bibitem{Collins:1984kg}
  J.~C.~Collins, D.~E.~Soper and G.~Sterman,
  Nucl.\ Phys.\ B {\bf 250} (1985) 199.

\bibitem{Sivers:1989cc}
  D.~W.~Sivers,
  Phys.\ Rev.\ D {\bf 41} (1990) 83,
  Phys.\ Rev.\ D {\bf 43}  (1991) 261.

\bibitem{Efremov:1992pe}
  A.~V.~Efremov, L.~Mankiewicz and N.~A.~Tornqvist,
  Phys.\ Lett.\ B {\bf 284} (1992) 394.

\bibitem{Collins:1992kk}
  J.~C.~Collins,
  Nucl.\ Phys.\ B {\bf 396} (1993) 161. 

\bibitem{Collins:1993kq}
  J.~C.~Collins, S.~F.~Heppelmann and G.~A.~Ladinsky,
  Nucl.\ Phys.\ B {\bf 420} (1994) 565.

\bibitem{Kotzinian:1994dv}
  A.~Kotzinian,
  Nucl.\ Phys.\  B {\bf 441} (1995) 234.

\bibitem{Mulders:1995dh}
  P.~J.~Mulders and R.~D.~Tangerman,
  Nucl.\ Phys.\ B {\bf 461} (1996) 197, 
  {\bf 484} (1997) 538E.

\bibitem{Boer:1997nt}
  D.~Boer and P.~J.~Mulders,
  Phys.\ Rev.\ D {\bf 57} (1998) 5780.

\bibitem{Boer:1997mf}
  D.~Boer, R.~Jakob and P.~J.~Mulders,
  Nucl.\ Phys.\ B {\bf 504} (1997) 345.

\bibitem{Boer:1999mm}
  D.~Boer,
  Phys.\ Rev.\  D {\bf 60} (1999) 014012.

\bibitem{Bacchetta:1999kz}
  A.~Bacchetta {\it et al.}, 
  Phys.\ Rev.\ Lett.\  {\bf 85} (2000) 712.

\bibitem{Collins:2002kn}
  J.~C.~Collins,
  Phys.\ Lett.\ B {\bf 536} (2002) 43. \  
  S.~J.~Brodsky, D.~S.~Hwang and I.~Schmidt,
  Phys.\ Lett.\ B {\bf 530} (2002) 99, 
  Nucl.\ Phys.\ B {\bf 642} (2002) 344.

\bibitem{Belitsky:2002sm}
  A.~V.~Belitsky, X.~Ji and F.~Yuan,
  Nucl.\ Phys.\ B {\bf 656} (2003) 165.
  X.~D.~Ji and F.~Yuan,
  Phys.\ Lett.\ B {\bf 543} (2002) 66.
  D.~Boer, P.~J.~Mulders and F.~Pijlman,
  Nucl.\ Phys.\ B {\bf 667} (2003) 201.
  I.~O.~Cherednikov and N.~G.~Stefanis,
  Phys.\ Rev.\  D {\bf 77}  (2008) 094001,
  Nucl.\ Phys.\  B {\bf 802} (2008) 146,
  Phys.\ Rev.\  D {\bf 80}  (2009) 054008.

\bibitem{Goeke:2005hb}
  K.~Goeke, A.~Metz and M.~Schlegel,
  Phys.\ Lett.\ B {\bf 618}  (2005) 90.

\bibitem{Bacchetta:2006tn}
  A.~Bacchetta, M.~Diehl, K.~Goeke, A.~Metz, P.~J.~Mulders and M.~Schlegel,
  JHEP {\bf 0702} (2007) 093.

\bibitem{Arnold:2008kf}
  S.~Arnold, A.~Metz and M.~Schlegel,
  Phys.\ Rev.\  D {\bf 79}  (2009) 034005.

\bibitem{Arneodo:1986cf}
  M.~Arneodo {\it et al.}  [European Muon Collaboration],
  Z.\ Phys.\ C {\bf 34} (1987) 277.

\bibitem{Airapetian:1999tv}
  A.~Airapetian {\it et al.}  [HERMES Collaboration],
  Phys.\ Rev.\ Lett.\  {\bf 84} (2000) 4047, 
  Phys.\ Rev.\ D {\bf 64} (2001) 097101, 
  Phys.\ Lett.\ B {\bf 562} (2003) 182, 
  Phys.\ Lett.\ B {\bf 622} (2005) 14.

\bibitem{Avakian:2003pk}
  H.~Avakian {\it et al.}  [CLAS Collaboration],
  Phys.\ Rev.\ D {\bf 69} (2004) 112004.\\
  A.~Airapetian {\it et al.}  [HERMES Collaboration],
  Phys.\ Lett.\  B {\bf 648} (2007) 164.

\bibitem{Airapetian:2004tw}
  A.~Airapetian {\it et al.}  [HERMES Collaboration],
  Phys.\ Rev.\ Lett.\  {\bf 94} (2005) 012002.

\bibitem{Alexakhin:2005iw}
  V.~Y.~Alexakhin {\it et al.}  [COMPASS Collaboration],
  Phys.\ Rev.\ Lett.\  {\bf 94} (2005) 202002.

\bibitem{Diefenthaler:2005gx}
  M.~Diefenthaler [HERMES Collaboration],
  AIP Conf.\ Proc.\  {\bf 792} (2005) 933.\\
  I.~M.~Gregor  [HERMES Collaboration],
  Acta Phys.\ Polon.\ B {\bf 36} (2005) 209.

\bibitem{Ageev:2006da}
  E.~S.~Ageev {\it et al.}  [COMPASS Collaboration],
  Nucl.\ Phys.\  B {\bf 765} (2007) 31.

\bibitem{Avakian:2005ps}
  H.~Avakian {\it et al.} 
  [CLAS Collaboration],
  AIP Conf.\ Proc.\  {\bf 792} (2005) 945.

\bibitem{Kotzinian:2007uv}
  A.~Kotzinian  [on behalf of the COMPASS Collaboration],
  arXiv:0705.2402 [hep-ex].

\bibitem{Osipenko:2008rv}
  M.~Osipenko {\it et al.}  [CLAS Collaboration],
  Phys.\ Rev.\  D {\bf 80}  (2009) 032004.

\bibitem{Giordano:2009hi}
  F.~Giordano, R.~Lamb  [HERMES Collaboration],
  AIP Conf.\ Proc.\  {\bf 1149} (2009) 423.\\
  A.~Bressan {\it et al.} [COMPASS Collaboration],
  arXiv:0907.5511 [hep-ex].

\bibitem{Gohn:2009}
  W.~Gohn, H.~Avakian, K.~Joo, and M.~Ungaro,
  AIP Conf.\ Proc.\  {\bf 1149} (2009) 461.

\bibitem{Abe:2005zx}
  K.~Abe {\it et al.}  [Belle Collaboration],
  Phys.\ Rev.\ Lett.\  {\bf 96} (2006) 232002.

\bibitem{Ogawa:2006bm}
  A.~Ogawa, M.~Grosse-Perdekamp, R.~Seidl and K.~Hasuko,
  arXiv:hep-ex/0607014.

\bibitem{Seidl:2008xc}
  R.~Seidl {\it et al.}  [Belle Collaboration],
  Phys.\ Rev.\  D {\bf 78} (2008) 032011.

\bibitem{Falciano:1986wk}
  S.~Falciano {\it et al.}  [NA10 Collaboration],
  Z.\ Phys.\  C {\bf 31} (1986) 513.

\bibitem{Guanziroli:1987rp}
  M.~Guanziroli {\it et al.}  [NA10 Collaboration],
  Z.\ Phys.\  C {\bf 37} (1988) 545.

\bibitem{Conway:1989fs}
  J.~S.~Conway {\it et al.},
  Phys.\ Rev.\  D {\bf 39} (1989) 92.

\bibitem{Zhu:2006gx}
  L.~Y.~Zhu {\it et al.},  
  Phys.\ Rev.\ Lett.\  {\bf 99} (2007) 082301, 
  {\bf 102} (2009) 182001.

\bibitem{Efremov:2004tp}
  A.~V.~Efremov {\it et al.},
  Phys.\ Lett.\  B {\bf 612} (2005) 233. 

\bibitem{Vogelsang:2005cs}
  W.~Vogelsang and F.~Yuan,
  Phys.\ Rev.\ D {\bf 72} (2005) 054028.

\bibitem{Anselmino:2005nn}
  M.~Anselmino  {\it et al.},
  Phys.\ Rev.\ D {\bf 71} (2005) 074006,
  Phys.\ Rev.\  D {\bf 75} (2007) 054032.

\bibitem{Collins:2005ie}
  J.~C.~Collins {\it et al.},
  Phys.\ Rev.\ D {\bf 73} (2006) 014021,
  Phys.\ Rev.\  D {\bf 73} (2006) 094023. 

\bibitem{Efremov:2006qm}
  A.~V.~Efremov, K.~Goeke and P.~Schweitzer,
  Phys.\ Rev.\  D {\bf 73} (2006) 094025.

\bibitem{Arnold:2008ap}
  S.~Arnold {\it et al.},
  arXiv:0805.2137 [hep-ph].

\bibitem{Anselmino:2008sga}
  M.~Anselmino {\it et al.},
  Eur.\ Phys.\ J.\  A {\bf 39} (2009) 89.

\bibitem{DeSanctis:2000fh}
  E.~De Sanctis, W.~D.~Nowak and K.~A.~Oganessian,
  Phys.\ Lett.\ B {\bf 483} (2000) 69.\\
  K.~A.~Oganessian {\it et al.}, 
  Nucl.\ Phys.\ A {\bf 689} (2001) 784.

\bibitem{Anselmino:2000mb}
  M.~Anselmino and F.~Murgia,
  Phys.\ Lett.\  B {\bf 483} (2000) 74.

\bibitem{Efremov:2001cz}
  A.~V.~Efremov, K.~Goeke and P.~Schweitzer,
  Phys.\ Lett.\ B {\bf 522} (2001) 37,
  {\bf 544} (2002) 389E,
  Eur.\ Phys.\ J.\ C {\bf 24} (2002) 407,
  Phys.\ Lett.\ B {\bf 568} (2003) 63,
  Phys.\ Rev.\  D {\bf 67} (2003) 114014;
  Eur.\ Phys.\ J.\ C {\bf 32} (2003) 337.


\bibitem{Ma:2002ns}
  B.~Q.~Ma, I.~Schmidt and J.~J.~Yang,
  Phys.\ Rev.\ D {\bf 66} (2002) 094001.

\bibitem{Bacchetta:2002tk}
  A.~Bacchetta, R.~Kundu, A.~Metz and P.~J.~Mulders,
  Phys.\ Rev.\  D {\bf 65} (2002) 094021.

\bibitem{Yuan:2003gu}
  F.~Yuan,
  Phys.\ Lett.\  B {\bf 589} (2004) 28. 

\bibitem{Gamberg:2003pz}
  L.~P.~Gamberg, D.~S.~Hwang and K.~A.~Oganessyan,
  Phys.\ Lett.\  B {\bf 584} (2004) 276.

\bibitem{Bacchetta:2004zf}
  A.~Bacchetta, P.~J.~Mulders and F.~Pijlman,
  Phys.\ Lett.\  B {\bf 595} (2004) 309. 

\bibitem{Metz:2004je}
  A.~Metz and M.~Schlegel,
  Eur.\ Phys.\ J.\  A {\bf 22} (2004) 489.

\bibitem{Metz:2004ya}
  A.~Metz and M.~Schlegel,
  Annalen Phys.\  {\bf 13} (2004) 699.

\bibitem{Afanasev:2006gw}
  A.~V.~Afanasev, C.~E.~Carlson,
  Phys.\ Rev.\  D {\bf 74} (2006) 114027, 
  arXiv:hep-ph/0308163.

\bibitem{Yuan:2003wk}
  F.~Yuan,
  Phys.\ Lett.\  B {\bf 575}, 45 (2003)
  [arXiv:hep-ph/0308157].

\bibitem{Gamberg:2006ru}
  L.~P.~Gamberg, D.~S.~Hwang, A.~Metz and M.~Schlegel,
  Phys.\ Lett.\  B {\bf 639} (2006) 508.

\bibitem{Gamberg:2007wm}
  L.~P.~Gamberg, G.~R.~Goldstein and M.~Schlegel,
  Phys.\ Rev.\  D {\bf 77} (2008) 094016. 

\bibitem{Jakob:1997wg}
  R.~Jakob, P.~J.~Mulders and J.~Rodrigues,
  Nucl.\ Phys.\  A {\bf 626} (1997) 937. 

\bibitem{Avakian:2008dz}
  H.~Avakian, A.~V.~Efremov, P.~Schweitzer and F.~Yuan,
  Phys.\ Rev.\  D {\bf 78} (2008) 114024, 
  arXiv:0808.3982 [hep-ph].

\bibitem{Pasquini:2008ax}
  B.~Pasquini, S.~Cazzaniga and S.~Boffi,
  Phys.\ Rev.\  D {\bf 78} (2008) 034025.

\bibitem{Efremov:2009ze}
  A.~V.~Efremov, P.~Schweitzer, O.~V.~Teryaev and P.~Zavada,
  Phys.\ Rev.\  D {\bf 80} (2009) 014021, 
  AIP Conf.\ Proc.\  {\bf 1149} (2009) 547. 

\bibitem{Meissner:2007rx}
  S.~Meissner, A.~Metz and K.~Goeke,
  Phys.\ Rev.\  D {\bf 76} (2007) 034002. 

\bibitem{She:2009jq}
  J.~She, J.~Zhu and B.~Q.~Ma,
  Phys.\ Rev.\  D {\bf 79} (2009) 054008. 

\bibitem{Bacchetta:2008af}
  A.~Bacchetta, F.~Conti and M.~Radici,
  Phys.\ Rev.\  D {\bf 78} (2008) 074010. 

\bibitem{work-in-progress}
  H.~Avakian, A.~V.~Efremov, P.~Schweitzer and F.~Yuan,
  forthcoming.

\bibitem{Courtoy:2008dn}
  A.~Courtoy {\it et al.}, 
  Phys.\ Rev.\  D {\bf 79} (2009) 074001, 
  Phys.\ Rev.\  D {\bf 78} (2008) 034002.

\bibitem{Boffi:2009sh}
  S.~Boffi, A.~V.~Efremov, B.~Pasquini and P.~Schweitzer,
  Phys.\ Rev.\  D {\bf 79} (2009) 094012,
  AIP Conf.\ Proc.\  {\bf 1149} (2009) 471, 
  arXiv:0910.1677 [hep-ph].

\bibitem{Kundu:2001pk}
  R.~Kundu and A.~Metz,
  Phys.\ Rev.\  D {\bf 65} (2002) 014009.\\
  M.~Schlegel and A.~Metz,
  arXiv:hep-ph/0406289.

\bibitem{Goeke:2003az}
  K.~Goeke, A.~Metz, P.~V.~Pobylitsa, M.~V.~Polyakov,
  Phys.\ Lett.\  B {\bf 567} (2003) 27.
 
\bibitem{Metz:2008ib}
  A.~Metz, P.~Schweitzer and T.~Teckentrup,
  Phys.\ Lett.\  B {\bf 680} (2009) 141. 

\bibitem{Teckentrup:2009xyz}
  T.~Teckentrup, A.~Metz, P.~Schweitzer, these proceedings 
  [arXiv:0910.2567 [hep-ph]].
 
\bibitem{Kotzinian:2006dw}
  A.~Kotzinian, B.~Parsamyan and A.~Prokudin,
  Phys.\ Rev.\  D {\bf 73} (2006) 114017.

\bibitem{Avakian:2007mv}
  H.~Avakian, A.~V.~Efremov, K.~Goeke, A.~Metz, P.~Schweitzer and T.~Teckentrup,
  Phys.\ Rev.\  D {\bf 77}  (2008) 014023.

\bibitem{Wandzura:1977qf}
  S.~Wandzura and F.~Wilczek,
  Phys.\ Lett.\  B {\bf 72} (1977) 195.

\bibitem{Jackson:1989ph}
   J.~D.~Jackson, G.~G.~Ross and R.~G.~Roberts,
   Phys.\ Lett.\  B {\bf 226} (1989) 159. 

\bibitem{Zavada:2002uz}
   P.~Zavada,
   Phys.\ Rev.\  D {\bf 67} (2003) 014019. 

\bibitem{D'Alesio:2009kv}
   U.~D'Alesio, E.~Leader and F.~Murgia,
   arXiv:0909.5650 [hep-ph].

\bibitem{Balla:1997hf}
  J.~Balla, M.~V.~Polyakov and C.~Weiss,
  Nucl.\ Phys.\  B {\bf 510} (1998) 327.

\bibitem{Accardi:2009au}
  A.~Accardi, A.~Bacchetta, W.~Melnitchouk and M.~Schlegel,
  arXiv:0907.2942 [hep-ph].\\
  A.~Accardi, A.~Bacchetta and M.~Schlegel,
  arXiv:0905.3118 [hep-ph].

\bibitem{Miller:2007ae}
  G.~A.~Miller,
  Phys.\ Rev.\  C {\bf 76} (2007) 065209. 

\bibitem{Jaffe:1991ra}
  R.~L.~Jaffe, X.~D.~Ji,
  Nucl.\ Phys.\  B {\bf 375} (1992) 527, 
  Phys.\ Rev.\ Lett.\  {\bf 67} (1991) 552.

\bibitem{Signal:1997ct}
  A.~I.~Signal, Nucl.\ Phys.\ B {\bf 497} (1997) 415.  

\bibitem{Efremov:2004tz}
  A.~V.~Efremov, O.~V.~Teryaev and P.~Zavada,
  Phys.\ Rev.\  D {\bf 70} (2004) 054018.
  P.~Zavada,
  arXiv:0908.2316 [hep-ph].

\bibitem{Schweitzer:2003uy}
  P.~Schweitzer,
  Phys.\ Rev.\  D {\bf 67} (2003) 114010.
  C.~Cebulla, J.~Ossmann, P.~Schweitzer and D.~Urbano,
  Acta Phys.\ Polon.\  B {\bf 39} (2008) 609.

\bibitem{Efremov:1983eb}
  A.~V.~Efremov and O.~V.~Teryaev,
  Sov.\ J.\ Nucl.\ Phys.\  {\bf 39}, 962 (1984)
  [Yad.\ Fiz.\  {\bf 39}, 1517 (1984)].

\bibitem{Teryaev:1995um}
  O.~V.~Teryaev,
  arXiv:hep-ph/0102296.


\bibitem{Anikin:2001ge}
  I.~V.~Anikin and O.~V.~Teryaev,
  Phys.\ Lett.\  B {\bf 509} (2001) 95.


\bibitem{Anikin:2009bf}
  I.~V.~Anikin, D.~Y.~Ivanov, B.~Pire, L.~Szymanowski and S.~Wallon,
  arXiv:0909.4090 [hep-ph].





\end{thebibliography}
\end{document}